\documentclass[12pt,preprint]{aastex}

\usepackage{amsmath}
\usepackage{listings}

\newcommand{\de}{{^{\circ}}}
\newcommand{\gs}{\mathrel{\raise0.35ex\hbox{$\scriptstyle >$}\kern-0.6em
\lower0.40ex\hbox{{$\scriptstyle \sim$}}}}
\newcommand{\ls}{\mathrel{\raise0.35ex\hbox{$\scriptstyle <$}\kern-0.6em
\lower0.40ex\hbox{{$\scriptstyle \sim$}}}}

\shorttitle{SPT }
\shortauthors{Stark}

\begin{document}

\title{ SPT-3G  secondary mirror geometry}

\author{Antony A. Stark}
\affil{Smithsonian Astrophysical Observatory, MS42, Cambridge MA 02138\\
\email{aas@cfa.harvard.edu}}

\begin{abstract}
SPT-3G is a detector system for the 10m diameter South Pole Telescope, 
comprising 16,000 millimeter-wave bolometers.  
It is used for a deep Cosmic
Microwave Background survey of the Southern sky.
This paper describes the geometry of the secondary mirror, which is a
section of a prolate spheroid, in several useful coordinate systems.
There is application to off-axis mirrors in general.
A geometric theorem is proven, relating to the Dragone condition:
the intersection of a prolate spheroid and any plane is an ellipse;
the lines connecting points on that ellipse to either focus compose a
right circular cone; the central axes of the two cones from the two foci
intersect outside the interior of the spheroid.

\end{abstract}

\bigskip

This paper describes the geometry of the SPT-3G
\citep{benson14,pan18} offset gregorian secondary mirror on the
South Pole Telescope \citep{carlstrom11},
with respect to a global coordinate system centered
on the primary mirror, and also the shape of the mirror in 
two coordinate systems local to the mirror itself (one referenced to the
central ray, and one referenced to the edge of the mirror).
This mirror has replaced the secondary mirror used in the
first two generations of SPT receivers.

The telescope is oriented pointed at the horizon,
toward the $+\hat x$
direction, so that the light from the star is moving in the $- x$
direction when intercepted by the primary mirror.  The $\hat y$ axis
points up, and the $\hat z$ axis is oriented according to the right-hand 
rule ($\hat x$ points right, $\hat y$ points up, and $\hat z$ points 
out of the page).  The origin is at the vertex of the primary
mirror.  This coordinate system is the one used by
Vertex.  It differs from the one
usually used in optics, by a rotation of   $90\de$
about the $\hat y$ axis:
$$
x_{\mathrm{zemax}} =  z, \,
y_{\mathrm{zemax}} =  y, \,
z_{\mathrm{zemax}} = -x \, .
$$

While using this memo, it may be useful to consult Nils Halverson's 
Optics Dimensional Control drawing: 
{\tt{OpticsDimControl.pdf}}, and A. Stark's program
{\tt{spt3gsecondary.c}} and its output, which can be
found under {\tt{http://spt.uchicago.edu/intweb/optics}}.

\section{The Central Ray}

The surface of the primary is defined by:
\begin{equation}
x = {{y^2 + z^2}\over{ 4\, f_{\mathrm{p}}}} \,
\end{equation}
where
$f_{\mathrm{p}} = 7000 \, \mathrm{mm}$ is the focal length of
the primary mirror.
Let $y_\mathrm{c} = 5300 \, \mathrm{mm}$.
The central ray starts at $(+\infty, \, y_\mathrm{c}, \, 0)$, and
intercepts the primary mirror at
$$
(y_\mathrm{c}^2 / 4 f_\mathrm{p}, \, y_\mathrm{c}, \, 0) = 
(1003.214286 \, \mathrm{mm}, \, 5300 \, \mathrm{mm}, \, 0)\, ,
$$ 
where it is reflected through a half-angle 
$$
i_{\mathrm{p}} = 
\mathrm{atan}({{y_\mathrm{c}}/{2 f_{\mathrm{p}}}}) =
20.735234\de \, ,
$$
and passes through the prime focus at
$$
\vec{F_1} = (f_\mathrm{p}, \, 0, \, 0) = 
(7000 \, \mathrm{mm}, \, 0, \, 0) \, .
$$
Between the primary and the secondary, the central ray satisfies
the equation:
\begin{equation}
y =  \alpha (x - f_\mathrm{p}) \,,
\end{equation}
where
$$
\alpha \equiv  {{4 f_\mathrm{p} y_\mathrm{c}}\over 
{y_\mathrm{c}^2 - 4 {f_{\mathrm{p}}^2}}}  = 
- \mathrm{tan}( 2 i_{\mathrm{p}}) 
= -0.8838068
\,,
$$
is the tangent of the angle from the
$+\hat x$ axis to the central ray between the primary and the secondary mirrors.

\begin{figure}[t!]
\centering
\vbox{
	\includegraphics[width=6.0in]{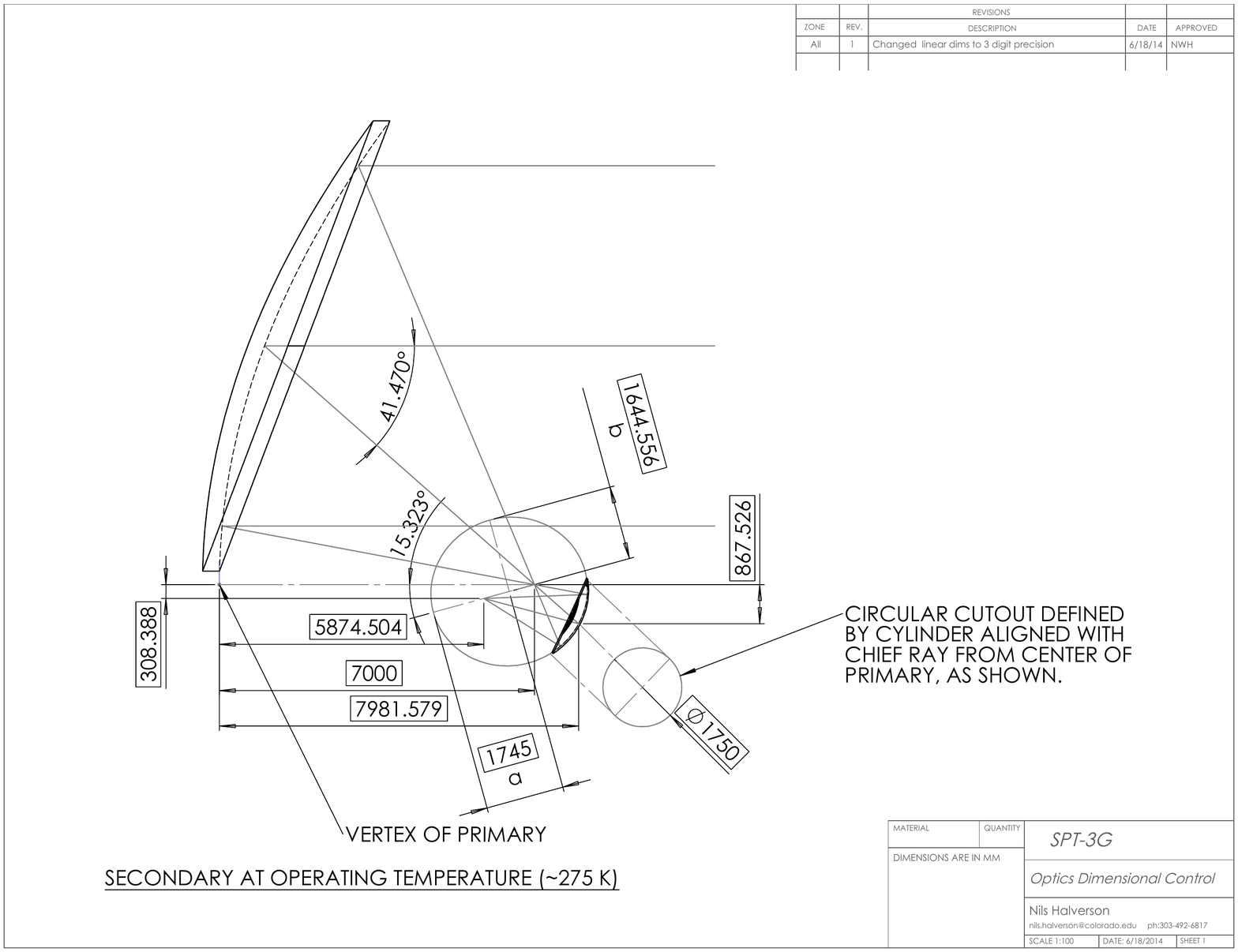}
	}
{\footnotesize
Dimensional drawing for SPT-3G secondary mirror from Nils Halverson.  Linear dimensions in millimeters.}
\vspace{-.1in}
\end{figure}

The SPT3G secondary is a piece of a prolate spheroid. 
Let $F_1$ be the focal point at the prime focus of the primary,
$F_2$ be the focus after reflection (the Gregorian focus), 
and $Q$ be the point where the
central ray strikes the surface and is reflected.
In the drawing {\em Optics Dimensional Control} 6/18/14 by Nils Halverson,
the spheroid is defined by a semi-major axis $a \, = \, 1745 \, \mathrm{mm}$
and semi-minor axis $b \, = \, 1644.556\,  \mathrm{mm}$.
This immediately gives the eccentricity $e \, = \, \sqrt{1 \, - \, (b/a)^2} 
\, = \, 0.334378$, conic constant $k \, = \, -e^2 \, = \, -0.111809$, 
and end-cap 
radius of curvature $R \, = \, a(1+k) \, = \, 1549.89366 \, \mathrm{mm}$.
The focal distance $f_0 \, = \, \sqrt{a^2 \, - \, b^2} \, = \, ae \, = \,
583.4896 \, \mathrm{mm}$ is half the length of the line segment $F_1 F_2$,
and the vertex 
distance $f_s \, = \, a - f_0 \, = \, 1161.5097 \, \mathrm{mm}$ is 
the length from vertex to focus.
The surface can be expressed as:
\begin{equation}
{{(x''+  f_0)^2}\over{a^2}}
 + {{{y''}^2}\over{b^2}}
 + {{{z''}^2}\over{b^2}} = 1 \, .
\end{equation}
The $(x'',\, y'', \, z'')$ coordinate system is defined with its origin
at the $F_1$ focus of the spheroid and the axis of the 
spheroid along the $\hat x''$ direction.
As shown in the {\em Optics Dimensional Control} diagram,
the secondary can be put in place by rotating by angle
$\theta_\mathrm{s} = 15.323\de$ 
around the $\hat z''$
axis then translating
in $\hat x$ by a distance
$f_{\mathrm{p}} = 7000 \, \mathrm{mm}$ to place the
$F_1$ focus coincident with the prime focus:
\begin{equation}
\begin{pmatrix}
x \\ y \\ z 
\end {pmatrix}
=
\begin{pmatrix}
{ \cos\,} \theta_\mathrm{s} &{\rm -sin\,} \theta_\mathrm{s}&0\\
{\rm sin\,} \theta_\mathrm{s}&{\rm cos\,} \theta_\mathrm{s}&0\\
0&0&1 
\end {pmatrix}
\begin{pmatrix}
x''\\ y''\\ z'' 
\end {pmatrix}
	+
\begin{pmatrix}
f_\mathrm{p} \\ 0\\ 0 
\end {pmatrix}
\, .
\end{equation}
The inverse transform is:
\begin{equation}
\begin{pmatrix}
x''\\ y''\\ z''
\end {pmatrix}
=
\begin{pmatrix}
{\rm cos\,} \theta_\mathrm{s} &{\rm sin\,} \theta_\mathrm{s}&0\\
{\rm -sin\,} \theta_\mathrm{s}&{\rm cos\,} \theta_\mathrm{s}&0\\
0&0&1 
\end {pmatrix}
\begin{pmatrix}
x-f_\mathrm{p}\\ y\\ z
\end {pmatrix}
\,.
\end{equation}
Substituting Equation 5 into Equation 3 yields an equation for the
secondary in $(x, \, y, \, z)$:
\begin{equation}
{{([x-f_\mathrm{p}]\,{\mathrm{cos}}\,\theta_\mathrm{s}
+ y\, {\mathrm{sin}}\,\theta_\mathrm{s} + f_0)^2}\over{a^2}}
+
{{([x-f_\mathrm{p}]\, {\mathrm{sin}}\,\theta_\mathrm{s}
- y\, {\mathrm{cos}}\,\theta_\mathrm{s} )^2}\over{b^2}}
 + {{{z}^2}\over{b^2}} = 1 \, .
\end{equation}
This equation can be solved for $y$ as a function of $x$ and $z$:
\begin{equation}
y = {{-\beta\gamma-\eta\lambda-\sqrt{(\beta\gamma+\eta\lambda)^2-
(\beta^2+\eta^2)(\gamma^2+\lambda^2+\mu^2 -1)}}\over{\beta^2+\eta^2}} \, ,
\end{equation}
where
\begin{eqnarray}
\beta & = & \mathrm{sin}\,\theta_\mathrm{s}/a \, , \nonumber\\
\gamma & = & ([x - f_\mathrm{p}]\mathrm{cos}\,\theta_\mathrm{s}
+f_0)/a \, , \nonumber\\
\eta & = & -\mathrm{cos}\,\theta_\mathrm{s}/b \, , \nonumber\\
\lambda & = &([x - f_\mathrm{p}]\mathrm{sin}\,\theta_\mathrm{s})/b \, , \nonumber\\
\mu & = & z/b  \, . \nonumber
\end{eqnarray}
Substituting Equation 2 and $z = 0$ into Equation 6 allows us to solve for
$x_\mathrm{sc}$, the $x$ coordinate of the
intersection of the central ray with the secondary:
\begin{equation}
x_\mathrm{sc} = {{-\rho\sigma+\sqrt{\rho^2\sigma^2-
(\rho^2+\tau^2)(\sigma^2-1)}}\over{\rho^2+\tau^2}} + f_\mathrm{p} 
= 7981.579 \, \mathrm{mm}\, ,
\end{equation}
where
\begin{eqnarray}
\rho & = & (\mathrm{cos}\,\theta_\mathrm{s}
+\alpha\,\mathrm{sin}\,\theta_\mathrm{s})/a \, , \nonumber\\
\sigma & = & f_0/a \,= \, e , \nonumber\\
\tau & = & (\mathrm{sin}\,\theta_\mathrm{s}
-\alpha\,\mathrm{cos}\,\theta_\mathrm{s})/b \, . \nonumber
\end{eqnarray}
Here we introduce a notation where a point in space is represented
by a vector extending from the origin of the $(x, \, y, \, z)$
coordinate system to that point.
Of course, these vectors can be
represented in other coordinate systems, even though they
depend on the $(x, \, y, \, z)$ coordinate system
for their definition. 
The central ray intersects the secondary at the point:
\begin{equation}
\vec{S} 
= (x_\mathrm{sc} , \, \alpha [x_\mathrm{sc} - f_\mathrm{p}] , \, 0) 
= (7981.579\,\mathrm{mm} , \, -867.52618,\mathrm{mm} , \, 0) 
\, ,
\end{equation}
and
the Gregorian focus is the point:
\begin{equation}
\vec{F_2} = 
(f_\mathrm{p} - 2 f_0 \, \mathrm{cos} \, \theta_\mathrm{s} \, , \,
- 2 f_0 \, \mathrm{sin} \, \theta_\mathrm{s} \, , \,
0 ) 
= (5874.5032\,\mathrm{mm} , \, -308.38675 \,\mathrm{mm} , \, 0) 
\, ,
\end{equation}
in the $(x, \, y, \, z)$ coordinate system.

\section{Dragone Relation}

The distance from the Gregorian focus to point $S$ on the secondary is:
\begin{equation}
f_2 =
\|\vec{F_2}-\vec{S}\| = 2180.00 \, \mathrm{mm}\, ,
\end{equation}
while the distance from the prime focus to point $S$ on the secondary is:
\begin{equation}
f_1 =
\|\vec{F_1}-\vec{S}\| = 1310.00\, \mathrm{mm}\, ,
\end{equation}
and the ratio of these distances is the magnification of the secondary:
\begin{equation}
M = -
{{
\|\vec{F_2}-\vec{S}\| 
}\over{
\|\vec{F_1}-\vec{S}\| 
}} 
= - {{f_2}\over{f_1}} = -1.6641236\, ,
\end{equation}
where by convention the magnification is negative for a Gregorian.
The angle of incidence at the secondary is:
\begin{equation}
i_\mathrm{s} = {{1}\over{2}}\,\mathrm{acos}
\left[{{(\vec{F_2}-\vec{S})\cdot
	(\vec{F_1}-\vec{S})}}
\over{
{\|\vec{F_2}-\vec{S}\|\|\vec{F_1}-\vec{S}\|}}\right] 
= 13.304411\de \, .
\end{equation}
The Dragone angle \citep{dragone82} is then:
\begin{equation}
\mathrm{tan}\,i_\mathrm{D} \equiv
(1-M)\mathrm{tan}\,i_\mathrm{s} +
M\mathrm{tan}\,i_\mathrm{p} = 0 \, .
\end{equation}
We see that it is zero, since $\theta_\mathrm{s}$ was chosen to yield that
result.

\section{The Secondary Mirror in Local Coordinates around the Central Ray}

Define a new coordinate system
$(x', y', z')$, where the origin is at $F_1$ like the 
$(x'', y'', z'')$ system, but rotated 
by angle $\theta_1$ about $\hat y''$ to place 
point $S$, the intersection of the central ray with
the mirror surface, on the positive $\hat z'$  
axis:
\begin{equation}
\begin{pmatrix}
x'\\ y'\\ z' 
\end{pmatrix}
= 
\begin{pmatrix}
{\rm cos\,} \theta_1 & 0 &{\rm sin\,} \theta_1\\
0&1&0\\
{-{\rm sin\,}} \theta_1 & 0 &{\rm cos\,} \theta_1
\end{pmatrix}
\begin{pmatrix}
x'' \\ y'' \\ z''  
\end{pmatrix}
\, ,
\end{equation}
and the inverse transform is
\begin{equation}
\begin{pmatrix}
x'' \\ y'' \\ z''
\end{pmatrix}
= 
\begin{pmatrix}
{\rm cos\,} \theta_1 & 0 &- {\rm sin\,} \theta_1\\
0&1&0\\
{\rm sin\,} \theta_1 & 0 &{\rm cos\,} \theta_1
\end{pmatrix}
\begin{pmatrix}
x'\\ y'\\ z' 
\end{pmatrix}
\, .
\end{equation}
Substituting Equation 17 into Equation 3 yields the 
equation of the spheroid in the 
$(x', y', z')$ system:
\begin{equation}
{{(x'\,{\mathrm{cos}}\,\theta_1
- z'\, {\mathrm{sin}}\,\theta_1   + f_0)^2}\over{a^2}}
+ {{{y'}^2}\over{b^2}}
+
{{(x'\, {\mathrm{sin}}\,\theta_1
+ z'\, {\mathrm{cos}}\,\theta_1 )^2}\over{b^2}}
= 1 \, .
\end{equation}
Setting $x' \, = \, y'\, = \, 0$, and solving the 
resulting quadratic for $z'$ gives the 
distance, $f_1$, between $F_1$ and $S$:
\begin{equation}
f_1 = {{b^2}\over{a}}\,
(1\,+\, e\,{\mathrm{sin}\,\theta_1})^{-1} \, .
\end{equation}
We see that $\theta_1$ is the complementry angle of the ``true anomaly"
of the point $S$.
We can solve for $\theta_1$:
\begin{eqnarray}
	{\mathrm{sin}\,\theta_1} \, &=& \,
\left({{b^2}\over{a f_1}} \, - \, 1 \right) \,e^{-1} \, , \\
\theta_1 \, &=& \, 33.206532\de \,.
\end{eqnarray}
Since $f_2$ is the distance from $Q$ to $F_2$, 
from the properties of ellipses we have
$a \, = \, {{1}\over{2}}(f_1 \, + \, f_2)$,
$b \, = \, \sqrt{{{1}\over{4}}(f_1+f_2)^2\,-\,f_0^2}$ ,
and we can express $\theta_1$ in terms of $f_0$, $f_1$, and
$f_2$ only:
\begin{equation}
{\mathrm{sin}\,\theta_1} \, = \,
{{f_2^2\,-\,f_1^2\,-\,4f_0^2}\over{4 f_0 f_1}} \,.
\end{equation}
This equation can also be derived by applying the law of cosines
to the triangle $F_1 C F_2$.

In the 
$(x', y', z')$ system, the coordinates of point $Q$ are
$(0, 0, f_1)$.  Substituting into Equation 17, the coordinates
of $Q$ in
$(x'', y'', z'') = (- \, f_1 {\mathrm{sin}\,\theta_1} , 0,
{f_1 \mathrm{cos}\,\theta_1})$ .
Solve Equation 3 for $z''$ as a function of $x''$, and differentiate 
to obtain the slope of the ellipse at point $C$:
\begin{equation}
{\mathrm{tan}\,\theta_2} \, = \,
{{{\mathrm{d}z''}}\over{{\mathrm{d}x''}}} \, = \,
-{{b}\over{a^2}} \, x'' \left(1 \, - \, {{{x'' \, }^2}\over{a^2}}\right)^{-1/2}
\, = \,
{{b}\over{a^2}} \, 
(f_1 {\mathrm{sin}\,\theta_1}\,+\, f_0)
\left[1 \, - \, {{{
(f_1 {\mathrm{sin}\,\theta_1}\,+\, f_0)
\, }^2}\over{a^2}}\right]^{-1/2}
\end{equation}

Let $i_\mathrm{s} \,= \, \theta_2 \, - \, \theta_1$.
Now we can define the 
$(x''', y''', z''')$ system, that has its origin at $S$ in the center of the mirror,
the $\hat x'''$-$\hat y'''$ plane tangent to the spheroid, and the $\hat z'''$
axis pointing into the mirror surface. 
\begin{equation}
\begin{pmatrix}
x'''\\ y'''\\ z'''
\end{pmatrix}
=
\begin{pmatrix}
{\rm cos\,} i_\mathrm{s} & 0 &{\rm sin\,} i_\mathrm{s}\\
0&1&0\\
- {\rm sin\,} i_\mathrm{s} & 0 &{\rm cos\,} i_\mathrm{s}
\end{pmatrix}
\begin{pmatrix}
x'  \\ y' \\ z' - f_1 
\end{pmatrix}
\, ,
\end{equation}
and the inverse transform is
\begin{equation}
\begin{pmatrix}
x' \\ y' \\ z'
\end{pmatrix}
=
\begin{pmatrix}
{\rm cos\,} i_\mathrm{s} & 0 & - {\rm sin\,} i_\mathrm{s}\\
0&1&0\\
{\rm sin\,} i_\mathrm{s} & 0 &{\rm cos\,} i_\mathrm{s}
\end{pmatrix}
\begin{pmatrix}
x'''\\ y'''\\ z'''
\end{pmatrix}
+
\begin{pmatrix}
0 \\ 0\\ f_1  
\end{pmatrix}
\, .
\end{equation}
Note that $i_\mathrm{s}$ is the angle of incidence at the secondary, since
the incident ray lies along the $\hat z'$ axis, and $\hat z'''$ is normal
to the surface.  Applying the law of cosines to the triangle
$F_1 S F_2$, we have:
\begin{equation}
	4 f_0^2 \, = \,  f_1^2 \, + \, f_2^2 \, - \, 2 f_1 f_2 {\mathrm{cos}\,(2 i_{\mathrm s})}
\, .
\end{equation}
We can now eliminate $a$, $b$, $e$, $f_0$, $\theta_1$, and $\theta_2$
in favor of $f_1$, $f_2$ and $i_\mathrm{s}$ in all of the preceeding equations.
The shape of the mirror can therefore be described in terms of
the two focal distances and the angle of incidence.
Combine Equation 25 with Equation 26, substitute into Equation 3,
and simplify: 
\begin{equation}
p z'''^2 \, + \, (q x''' \,+\, 2\, r)z''' \, + \, y'''^2 \, + \, s x'''^2 \, = \, 0 \, ,
\end{equation}
where
\begin{eqnarray}
p \, &=& 1\, - \,e^2\,{\mathrm{sin}}^2 \theta_2 \, = \, 0.94114858\\
q \, &=& \, e^2 \, {\mathrm{sin}} \, (2 \, \theta_2) \, = \, 0.11165324\\
	r \, &=& \, f_1( {\mathrm{cos}} \, i_\mathrm{s} + e \,{\mathrm{sin}} \, \theta_2 ) 
	\, = \, f_2( {\mathrm{cos}} \, i_\mathrm{s} - e \,{\mathrm{sin}} \, \theta_2 )  \, = \, 1592.6382 \, \mathrm{mm}\\
s \, &=& 1\, - \,e^2 \,{\mathrm{cos}}^2 \theta_2 \, = \, 0.94704272
\end {eqnarray}
and
\begin{eqnarray}
f_0 \, &=& \,  {{1}\over{2}}\, \sqrt{f_1^2 \, + \, f_2^2 \, 
- \, 2 f_1 f_2 \,{\mathrm{cos}\,(2 i_\mathrm{s})}} \, = \, 583.49067 \, \mathrm{mm}\\
\theta_2 \, &=& \, i_\mathrm{s} \,+\, {\mathrm{arcsin}}\left(
{{f_2^2\,-\,f_1^2\,-\,4f_0^2}\over{4 f_0 f_1}} \right) \, = \, 46.51091\de \\
e \, &=& \, {{2 f_0}\over{f_1\,+\,f_2}} \, = \, 0.334378  \, ,
\end {eqnarray}
yielding the equation of the mirror surface sag in $(x''', y''', z''')$ coordinates:
\begin{equation}
z''' \, = \, {{1}\over{2p}}\left[ \sqrt{
(2\, r \, + \, q x''')^2 \, - \, 4 p \, (y'''^2 \, + \, s x'''^2)} 
\, - \, (2\, r \, + \, qx''' ) \, \right] \, .
\end{equation}
This is the shape of the secondary mirror in a coordinate system centered
on the intersection of the central ray with the surface, where the
surface sag ($z'''$) is normal to the surface at that point.
The value of $r$ 
has dimensions of length,
and it is always positive (either of the expressions in
equation 30 can be used, depending on the sign of $\theta_2$). 
It is the radius
of curvature in the $\hat y'''$-$\hat z'''$ plane, 
but it is not the radius of curvature seen by the beam, 
as will be shown in Equation 36.
The values of $p$ and $s$ are dimensionless and always between 0 and 1.
The value of $q$ is dimensionless and
always between $-1$ and 1; its sign is opposite 
that of the $x''$
coordinate of the point $S$.
Note that $z'''$ is everywhere negative, since $p$ and $s$ are
positive in Equation 35.  The $\hat z'''$ direction points into the
mirror, and the $\hat x'''$-$\hat y'''$ plane is tangent to the
mirror at point $S$, the origin of the   
$(x, y, z)$ system.  The direction of $\hat x'''$ is such that its
dot product with the
vector from $F_1$ to $F_2$ is positive.

Since we know the two focal distances $f_1$ and $f_2$, the
thin lens formula gives the paraxial focal length $f$:
\begin{equation}
{{1}\over{f}} \, = \, {{1}\over{f_1}}
\, + \, {{1}\over{f_2}} \, = \,
{{2 \, \mathrm{cos}\, i_\mathrm{s}} \over {r}} \, ,
\end{equation}
using Equation 30 to substitute for $f_1$ and $f_2$.
So $f\, = \, r / (2 \, {\mathrm{cos}\, i_\mathrm{s}}) \, 
= \, 818.28089 \, \mathrm{mm}$, and
the central radius of curvature is $r / {\mathrm{cos}\, i_\mathrm{s}} \, = \, 
1636.5618 \, \mathrm{mm}$.

\section{Defining the Edge of the Secondary}

Unlike the first SPT receiver and SPTpol, the edge of the secondary
is not a stop of the optical system, and so the secondary need only
be big enough not to significantly vingnette; the exact figure of
the edge doesn't matter optically.
For convenience in manufacturing, however, we'd like the edge of the
secondary to lie in a plane.  This can be accomplished by defining
the edge as the intersection of the secondary spheroid with a
right circular cone whose vertex is at the Gregorian focus
$\vec{F_2}$ and whose axis is
coincident with the central ray
$\vec{F_2}-\vec{Q}$.
The half-angle of the cone is defined to
be $\phi_2 \, = \, 23.5993\de$,
and the angle between the rotation axis of the
cone and the line between the focii is
$\psi_2 \, = \, -30.1844\de$.
These two angles, together with the major and minor
axes of the spheroid ($a$ and $b$), fully define the
mirror.  The analysis is done in
the Appendix below, and the results presented in Table 1.
The calculations in the Appendix show that there are equivalent
ways of defining the edge of the secondary: a right circular
cone whose vertex is the prime focus, $F_1$, with
half angle
$\phi_1 \, = \, 38.8953\de$
and tilt
$\psi_1 \, = \, -52.0529\de$
also intersects the
spheroidal surface of the secondary in the same ellipse
as the cone from $F_2$, although the axis of that cone does not
lie on the central ray of the optics.

As a practical matter, the edge of the secondary mirror is
chosen to be circular, the result of intersecting the prolate
spheroid with a cylinder.  The edge does not, therefore, lie in
a plane.  The edge of the primary is not illuminated by the detector
optics, so the precise geometry does not matter.

\section{Mirror Coordinates Relative to the Mirror Edge}

Given the definition of the edge of the secondary, we can
transform into
a coordinate system centered on the ellipse that
is the edge of the secondary and tilted so
that the plane of the mirror edge is the $x''''-y''''$ plane at
$z'''' \, = \, 0$: 
\begin{equation}
\begin{pmatrix}
x''''\\ y''''\\ z''''
\end{pmatrix}
=
\begin{pmatrix}
{\rm cos\,} \theta_\mathrm{e} &{\rm sin\,} \theta_\mathrm{e}&0\\
0&0&-1\\
{-\rm sin\,} \theta_\mathrm{e}&{\rm cos\,} \theta_\mathrm{e}&0
\end{pmatrix}
\begin{pmatrix}
x-c\\ y \\ z
\end{pmatrix}
+
\begin{pmatrix}
d \\ 0 \\ 0
\end{pmatrix}
\,.
\end{equation}
The inverse transform is:
\begin{equation}
\begin{pmatrix}
x\\ y\\ z
\end{pmatrix}
=
\begin{pmatrix}
{\rm cos\,} \theta_\mathrm{e} &0&-{\rm sin\,} \theta_\mathrm{e}\\
{\rm sin\,} \theta_\mathrm{e}&0&{\rm cos\,} \theta_\mathrm{e}\\
0&-1&0
\end{pmatrix}
\begin{pmatrix}
x'''' - d\\ y''''\\ z''''
\end{pmatrix}
+
\begin{pmatrix}
c \\ 0 \\ 0
\end{pmatrix}
\, ,
\end{equation}
where in this case $(x, \, y, \, z)$ is the coordinate 
system of Equation 41 and not the global coordinate system.
The offsets $c$ and $d$ are given in Table 1,
and $\theta_e \, = \, 47.9795\de$ is the 
angle $\theta $ defined in the Appendix
($\theta_e$, the slope of the mirror edge, 
differs from $\theta_2$ in Equation 33, the slope of the mirror
surface at the central ray).
This is the coordinate system needed to cut the mirror surface.
In this coordinate system, the equation of the secondary mirror can be
expressed as:
\begin{equation}
z'''' = {{P x''''+Q+a \sqrt{J y''''^2+K x''''^2+ Lx'''' +N}}\over{J}} \,
\end{equation}
where
\begin{eqnarray}
	c & = & (f_0 \, \cos \psi_2 \, - \, a \cos \phi_2)/\sin \, \psi_2 \, \tan \theta_\mathrm{e} \, = \, 1961.81 \, \mathrm{mm} \nonumber \\
	d & = & b^2 \, c \cos \theta_\mathrm{e}/(a^2 \, \sin^2 \theta_\mathrm{e}
	\, + \, b^2 \cos^2 \theta_\mathrm{e}) 
	\, = \, 1227.92 \, \mathrm{mm}\nonumber \\
	J & = & -f_0^2\, \mathrm{cos}^2 \theta_\mathrm{e}\,-\, b^2 \, 
	= \, -2857121.6 \, \mathrm{mm}^2 \nonumber \\
	K & = & -b^2 \, = \, -2704564.4 \, \mathrm{mm}^2 \nonumber \\
L & = & 2 \, b^2\, [ d \, - \, 
          c \, \mathrm{cos}\,(\theta_\mathrm{e})] \, = \,
	  -4.614588\,\times\,10^{8} \, \mathrm{mm}^3 \nonumber \\
N & = &  b^2\{b^2+f_0^2\,\mathrm{cos}^2\,\theta_\mathrm{e} \,-\,
	 [d \, - \, c \, \mathrm{cos}\,\theta_\mathrm{e}]^2\} \, = \,
	 7.7075858 \, \times\,10^{12} \, \mathrm{mm}^4 \nonumber \\
P & = & f_0^2 \, \mathrm{cos} \, \theta_\mathrm{e}\,
	 \mathrm{sin} \, \theta_\mathrm{e}
         \, = \, 169310.4\, \mathrm{mm}^2 \nonumber \\
 Q & = &  -[f_0^2 \, d \, \cos \, \theta_\mathrm{e} + b^2\,c ] \,
	 \mathrm{sin}\, \theta_\mathrm{e} \, = \,
	 -4.1496374 \, \times \, 10^9 \mathrm{mm}^3 \nonumber
\, .
\end{eqnarray}
The $\hat x''''$ direction is up, and the top of the mirror is
$(x'''', \, y'''', \, z'''') \, = \, (869.665\, \mathrm{mm}, \, 0, \, 0)$.
The point
$(x'''', \, y'''', \, z'''') \, = \, (0, \, 847.599, \, 0)$
is also on the mirror edge.
The point of maximum depth in the secondary mirror is not below the 
center of the mirror (that is, $x''''=0$, $y''''=0$), 
but is slightly displaced: 
\begin{eqnarray}
x''''_\mathrm{min}& = &{{a^2KL-L P^2-P\sqrt{(L^2-4KN)P^2+4a^2K^2N-a^2KL^2}}\over
                      {2(KP^2-a^2K^2)}}\\
                 &  & = 14.2408\, \mathrm{mm} \nonumber \\
y''''_\mathrm{min}& = & 0 \nonumber \\
z''''_\mathrm{min}& = & -243.286\, \mathrm{mm} \nonumber 
\end{eqnarray}
is the lowest point on the mirror surface, about a half inch above the center.

\newpage

\section{Appendix: Theorems about Spheroids, Cones, and Planes}

The coordinate systems and variables defined in this appendix
are independent of the body of this memo above.
Consider a prolate spheroid:
\begin{equation}
{{x^2}\over{a^2}}
+ {{y^2}\over{b^2}}
 + {{z^2}\over{b^2}} = 1 \, ,
\end{equation}
where
$a \, > \, b \, > \, 0$.
In the $(x, \, y, \, z)$ coordinate system, the
focii are $F_1 \, = \, (+ f_0 , \, 0, \, 0)$
and $F_2 \, = \, (- f_0 , \, 0, \, 0)$,
where $f_0 \, \equiv \,  \sqrt{a^2 \, - \, b^2}$.
Cut the spheroid with an arbitrary plane, 
$y \, = \, (x - c)\, {\mathrm{tan}\,}\theta $, where 
$ - 90\de \, < \, \theta \, < \, 90\de $.
By symmetry, this equation can describe any cutting of the
spheroid except for the special cases 
of a plane parallel to the axis of the spheroid
($\theta = 0\de$ exactly) or a plane perpendicular to the axis of the spheroid
($\theta = 90\de$ exactly).
These singular cases can easily be treated separately.
Transform so that the plane becomes the $\hat{u} - \hat{z}$ plane,
and translate in the $\hat{u}$ direction by an offset, $d$,
to be determined below:
\begin{equation}
\begin{pmatrix}
x\\ y\\ z
\end{pmatrix}
=
\begin{pmatrix}
{\rm cos\,} \theta &{\rm -sin\,} \theta&0\\
{\rm sin\,} \theta&{\rm cos\,} \theta&0\\
0&0&1
\end{pmatrix}
\begin{pmatrix}
u - d\\ v\\ z
\end{pmatrix}
	+
\begin{pmatrix}
c \\ 0\\ 0
\end{pmatrix}
\, ,
\end{equation}
and the inverse transform is
\begin{equation}
\begin{pmatrix}
u\\ v\\ z
\end{pmatrix}
=
\begin{pmatrix}
{\rm cos\,} \theta &{\rm  sin\,} \theta&0\\
{\rm - sin\,} \theta&{\rm cos\,} \theta&0\\
0&0&1
\end{pmatrix}
\begin{pmatrix}
x - c \\ y\\ z
\end{pmatrix}
	+
\begin{pmatrix}
d \\ 0\\ 0 
\end{pmatrix}
\, .
\end{equation}
Substitute Equation 42 into Equation 41, and set $v \, = \, 0$,
(which we see from Equation 43 gives 
$y \, = \, (x - c)\, {\mathrm{tan}\,}\theta $) yielding:
\begin{equation}
{{[(u-d)\,{\rm cos\,}\theta + c\,]^2}\over{a^2}}
+ {{(u-d)^2\,{\rm sin}^2 \theta}\over{b^2}}
 + {{z^2}\over{b^2}} = 1 \, .
\end{equation}
Let
\begin{equation}
	g^2 \, \equiv \, a^2 \,{\rm sin}^2 \theta \, + \, b^2 \,{\rm cos}^2 \theta
\end{equation}
and
\begin{equation}
	\gamma \, \equiv \, 1 \, - \, {{c^2 \,{\rm sin}^2 \theta} \over {g^2} }\, .
\end{equation}
Then, choosing the arbitrary translation in the $\hat u$ direction to be:
\begin{equation}
	d \, \equiv  \, {{b^2 c \,{\rm cos}\, \theta} \over {g^2} }\, ,
\end{equation}
Equation 44 can be written as
\begin{equation}
{{u^2}\over{\alpha^2}}
 + {{z^2}\over{\beta^2}} = 1 \, ,
\end{equation}
with
\begin{equation}
\alpha^2 \, \equiv \, {{a^2 b^2}\over{g^2}} \gamma
\end{equation}
and
\begin{equation}
\beta^2 \, \equiv \, b^2 \, \gamma \, .
\end{equation}
Equation 48 shows that the intersection of the spheroid and plane is
an ellipse, provided $\gamma \, > \, 0$ 
or $c^2 \, < \, a^2 \, + \, b^2 \, {\rm cot}^2 \theta $
(i.e. the plane actually intersects the spheroid and doesn't miss it).
{\em The intersection of a prolate spheroid and an arbitrary plane is
an ellipse.} (More generally, any closed figure that results from the intersection of a plane with an ellipsoid, paraboloid, or hyperboloid is an ellipse.)

In the $(u, \, v, \, z)$ coordinate system, the center of the 
ellipse is $C \, = \, (0, \, 0, \, 0)$, and the extrema of the ellipse are
$A \, = \, (\pm \, a b g^{-1}\sqrt{ \gamma} , \,  0, \, 0)$, 
and $B \, = \, (0, \, 0, \, \pm\,b \sqrt{ \gamma })$; 
the focii are $F_1 = ( [ f_0 - c ] {\rm cos} \, \theta + d, \, - [ f_0 - c] {\rm sin} \, \theta , \, 0)$
and $F_2 = ( - [  f_0 + c ] {\rm cos} \, \theta + d, \, - [f_0 + c] {\rm sin} \, \theta , \, 0) \, .$ 

In the $(x, \, y, \, z)$ coordinate system, the center of the 
ellipse is $C = (- d \, {\rm cos}\,\theta \, + \, c , \, -d \, {\rm sin}\, \theta, \, 0)$, 
and the extrema of the ellipse are
\begin{equation}
	A \, = \, ([\pm \, a b g^{-1}\sqrt{ \gamma} \, - \, d\, ] \, {\rm cos}\,\theta \, + \, c, \,  
	[\pm \, a b g^{-1}\sqrt{\gamma}\,  - \, d\, ]\, {\rm sin}\,\theta , 
 \, 0) \, , 
\end{equation}
and
\begin{equation}
B \, = \, ( - d \, {\rm cos}\,\theta \, + \, c , 
       \, -d \, {\rm sin}\, \theta, \, 
 \pm\,b \sqrt{ \gamma }) \, . 
\end{equation}

Consider
a right circular cone whose vertex is $F_1$, and whose
axis passes through the point $Q_1$. The 
points on cone satisfy the equation:
\begin{equation}
(\vec{P}-\vec{F_1})\cdot
(\vec{Q_1}-\vec{F_1})=
\|\vec{P}-\vec{F_1}\|\|\vec{Q_1}-\vec{F_1}\|\,\mathrm{cos}\,\phi_1  \, ,
\end{equation}
where $P$ is any point on the cone, $\phi_1$ is the opening
half-angle of the cone, and $Q_1$ is a point in the $x - y$ plane that
defines the axis of the cone.  Without loss of generality,
we take $ 0 \, < \, \phi_1 \, < \, 90\de $, so ${\rm cos} \, \phi_1 \, > 0$.

Define a coordinate system translated so that the origin is at $F_1$:
$x' \, = \, x - f_0$, $y' = y$, $z' = z$, and let the angle $\psi_1$
be the angle between the unit vector in the $x'$ direction and
the axis of the cone, $\vec{Q_1}-\vec{F_1}$:
\begin{equation}
\hat x' \cdot
(\vec{Q_1}-\vec{F_1})=
\|\vec{Q_1}-\vec{F_1}\|\,\mathrm{cos}\, \psi_1   \, .
\end{equation}
Without loss of generality, we take
we take $ - 90\de \, < \, \psi_1 \, < \, 90\de $, so ${\rm cos} \, \psi_1 \, > 0$.

Then Equation 53 becomes:
\begin{equation}
	x' \, {\rm cos}\, \psi_1 + y' \, {\rm sin} \, \psi_1 \, = \,
	\sqrt{x'^2 \, + \, y'^2 \, + \, z'^2} \, \, {\rm cos} \, \phi_1 \, .
\end{equation}
Solve Equation 55 for $z'^2$, substitute into Equation 41, and solve for $y$
to yield:
\begin{equation}
	y = {{(\pm f_0 \, {\rm cos}\, \phi_1 - a \, {\rm cos}\, \psi_1) \, x 
	+ a\, f_0 {\rm cos}\, \psi_1 \,  \mp
a^2 \, {\rm cos}\, \phi_1}\over{a \, {\rm sin}\,\psi_1}} \,.
\end{equation}
We see that
\begin{equation}
	{\rm tan}\,\theta \, = \,{{\pm \,  e \, {\rm cos}\, \phi_1 - 
	 {\rm cos}\, \psi_1 } \over { {\rm sin} \, \psi_1}} \, ,
\end{equation}
and
\begin{equation}
	c \, = \, 
	{{\pm \,  a \, {\rm cos}\, \phi_1 \, - \,
	 f_0 \, {\rm cos}\, \psi_1 }
        \over
        {{\rm sin}\, \psi_1 \, {\rm tan} \, \theta}} \, ,
\end{equation}
where $e \, \equiv \, f_0/a$ is the eccentricity of the spheroid,
describe the two planes 
\begin{equation}
y \, = \, (x - c)\, {\mathrm{tan}\,}\theta 
\end{equation}
whose intersection with the spheroid
yields the same ellipses as the intersection of the spheroid
with the two nappes of the cone 
that has a vertex at $F_1$, is tilted by angle $\psi_1$
with respect to the $\hat x$ axis, and has a half opening
angle of $\phi_1$.
Choosing the $+$ signs in Equations 57 and 58 yields the nappe
of the cone to the left, 
where the points on its axis satisfy $x \, < \, f_0$.
Note that $\psi_1$ and $\theta$ have opposite signs for the nappe
to the right, and usually have opposite signs for the nappe on the
left unless the spheroid is sufficiently eccentric 
($e \, > \, {\cos \, \psi_1}/{\cos \, \phi_1}$).
{\em The intersection of a prolate spheroid with a right circular
cone, whose vertex is one of the focii of the spheroid,
is a planar figure, specifically an ellipse.}
That is not true in general of a cylinder whose axis passes
through the focus --- the intersection of that cylinder with
the spheroid will not lie in a plane --- because the 
intersection of the cylinder
with a circular cone sharing the same axis is a circle, not
an ellipse, and we know that the intersection between the
cone and the spheroid is in general elliptical.

By symmetry, this result applies equally well to a cone whose
vertex is at $F_2$, but if we define the tilt of such a cone
with respect to the positive $\hat x$ axis:
\begin{equation}
\hat x'' \cdot
(\vec{Q_2}-\vec{F_2})=
\|\vec{Q_2}-\vec{F_2}\|\,\mathrm{cos}\, \psi_2   \, .
\end{equation}
where $x'' \, = \, x \, + \, f_0$, that breaks the symmetry
and we get slightly different equations for planes:
\begin{equation}
	{\rm tan}\,\theta \, = \,{{\mp \,  e \, {\rm cos}\, \phi_2 - 
	 {\rm cos}\, \psi_2 } \over { {\rm sin} \, \psi_2}} \, ,
\end{equation}
and
\begin{equation}
	c \, = \, 
	{{\pm \,  a \, {\rm cos}\, \phi_2 \, + \,
	 f_0 \, {\rm cos}\, \psi_2 }
        \over
        {{\rm sin}\, \psi_2 \, {\rm tan} \, \theta}} \, ,
\end{equation}
where like before, we take
$ 0 \, < \, \phi_2 \, < \, 90\de $ and
$ -90\de \, < \, \psi_2 \, < \, 90\de $.
Choosing the upper signs in Equations 61 and 62 yields the nappe
to the left (points on the axis satisfy $x \, < \, -f_0$).

Equations 57 and 58 can be solved to yield $\phi_1$ and $\psi_1$ as
a function of $\theta$ and $c$:
\begin{equation}
	{\rm tan}\,\psi_1 \, = \,{{b^2 \, {\rm cot}\, \theta}  
	\over {c \, f_0 \, - \, a^2 }} \, ,
\end{equation}
and
\begin{equation}
	\cos \phi_1 \, = \, 
	\left| {{\tan \, \theta \, \sin \, \psi_1 \, + \, \cos \, \psi_1}
	\over{e}} \right|
	\, ,
\end{equation}
while Equations 61 and 62 can be similarly be inverted to yield:
\begin{equation}
	{\rm tan}\,\psi_2 \, = \,{{- b^2 \, {\rm cot}\, \theta}  
	\over {c \, f_0 \, + \, a^2 }} \, ,
\end{equation}
and
\begin{equation}
	\cos \phi_2 \, = \, 
	\left| {{\tan \, \theta \, \sin \, \psi_2 \, + \, \cos \, \psi_2}
	\over{e}} \right|
	\, .
\end{equation}
For any plane defined by $\theta$ and $c$ that cuts the spheroid, 
there are right circular cones
from each focus defined by $\phi_1$, $\psi_1$, $\phi_2$, and
$\psi_2$ that generate the same ellipse as the plane.
{\em Any arbitrary plane cutting a prolate spheroid results in
an ellipse.  The set of lines connecting points on that ellipse
to the focii of the spheroid comprise two right circular cones
whose vertices lie on the focii.}

The points $Q_1$ and $Q_2$ 
cannot in general be coincident with the center of the
ellipse, point $C$.  When an ellipse
is generated by cutting a cone with a plane, the center of the
ellipse does not fall on the axis of the cone, except in the
singular case where the ellipse is a circle.
The location of $Q_1$ could be chosen to lie anywhere on
the line
$y \, = \, (x - f_0)\, {\mathrm{tan}\,}\psi_1 $, 
while the location of $Q_2$ could be chosen to lie anywhere on
$y \, = \, (x + f_0)\, {\mathrm{tan}\,}\psi_2 $. 
From Equations 63 and 65, we see that these two lines intersect
at the point
\begin{equation}
	Q_0 \, = \, \left({{a^2}\over{c}}, \,
	{{-b^2}\over{c \, \tan \, \theta}},\,0\right)
	\, ,
\end{equation}
and this point can serve as both $Q_1$ and $Q_2$, but in general
this point does not lie on the spheroid.
$Q_0$ is, of course, the intersection of the axes of the two cones,
and $c$ is the intercept of the plane with the $\hat x$ axis and
$- c  \tan \theta$ is the intercept of the plane with the $\hat y$
axis. $Q_0$ lies on the spheroid if 
$c^2 \, = \, a^2 \, + \, b^2\,\cot^2 \theta$, but that implies
that the plane is tangent to the spheroid,
$\gamma \, = \, 0$ in Equation 46, and the ellipse has shrunk to
a point --- not interesting
or useful.  The singular on-axis case,
where $\psi_1 \, = \, \psi_2 \, = \, 0$ and the ellipse is
actually a circle,
allows $Q_0 \, = (a, \, 0, \, 0)$.
In the general case, point $Q_0$ lies outside the spheroid.

The points $Q_1$ and $Q_2$ could also lie in the plane of the
ellipse or on the surface of the spheroid.
The intersections of the cone  axes with 
the line 
$y \, = \, (x - c)\, {\mathrm{tan}\,}\theta $, 
in the plane of the ellipse,
in $(x,\,y,\,z)$ coordinates, are:
\begin{equation}
Q_{1e} \, = \left({{c \, {\rm tan} \, \theta - f_0 \, {\rm tan} \, \psi_1}\over
{{\rm tan} \, \theta \, -  \, {\rm tan} \, \psi_1}}, \,
{{[c \, - \, f_0] \, {\rm tan} \, \theta \tan \, \psi_1 }\over
{{\rm tan} \, \theta \, -  \, {\rm tan} \, \psi_1}}\, , \, 0 \right) \, ,
\end{equation}
\begin{equation}
Q_{2e} \, = \left({{c \, {\rm tan} \, \theta + f_0 \, {\rm tan} \, \psi_2}\over
{{\rm tan} \, \theta \, -  \, {\rm tan} \, \psi_2}}, \,
{{[c \, + \, f_0] \, {\rm tan} \, \theta \tan \, \psi_2 }\over
{{\rm tan} \, \theta \, -  \, {\rm tan} \, \psi_2}}\, , \, 0 \right) \, .
\end{equation}
The intersections of the cone axes
with the spheroid are:
\begin{align}
	Q_{1s} \, = \,&  \left(
{{ a^2 f_0\,\tan ^2  \psi_1
\pm a b^2 \sec \psi_1}
 \over{a^2\,\tan ^2\, \psi_1+b^2}}\, , 
 {{
 - b^2 f_0\,
 \pm a b^2 \sec \psi_1}
 \over{a^2\,\tan \psi_1+b^2 \, \cot \psi_1
 }} ,\, 
 0
 \right) \,,
\end{align}
\begin{align}
	Q_{2s} \, = \,&  \left(
{{ - a^2 f_0\,\tan ^2  \psi_2
\pm a b^2 \sec \psi_2}
 \over{a^2\,\tan ^2\, \psi_2+b^2}}\, , 
 {{
 + b^2 f_0\,
 \pm a b^2 \sec \psi_2}
 \over{a^2\,\tan \psi_2+b^2\,\cot\psi_2
 }} , \, 
 0
 \right) \,.
\end{align}
Choose the $+$ sign for the nappe to the right.

If the cone represents a bundle of light rays coming from $F_2$ and
striking the spheroidal mirror, then $\vec Q_2 - \vec F_2$ is the
central ray, and it strikes the mirror at point $Q_{2s}$.
The light rays in the cone will strike the mirror and
reflect, converging in a different cone onto $F_1$.
{\em An ellipse that is the intersection of a spheroid and a
right circular cone whose vertex is one focus of the spheroid
can also be described as the intersection of the spheroid with
a different right circular cone whose vertex is the other focus.}
This second cone will have a different
opening angle $\phi_1$ and tilt $\psi_1$.  Also, the axes of the two cones
do not, in general, intersect at 
the spheroidal surface, so the 
points $Q_{1s}$ and $Q_{2s}$
will be different.
The central beam coming from $F_2$ will strike the mirror and be
reflected {\em before} it gets to $Q_0$.
This is important for optics, because if the axis of one cone is
the central ray of the beam, that ray will not reflect onto the
axis of the other cone, causing a skewness in the beam that can
only be corrected by an appropriate choice of angles in
subsequent reflections, i.e. the Dragone condition.

\begin{table}[!htbp]
\begin{center}
\begin{tabular}{l l r r r r}
\hline\hline
Variable & & & & & Coordinate\\
	 & & & & & System \\ [0.5ex]
\hline
$a$ &  semimajor axis of spheroid &1745.000 & & & \\
$b$ &  semiminor axis of spheroid &1644.556 & & & \\
$f_0$& focal distance of spheroid&  583.490 & & & \\
$\theta$ & angle of plane&  $47.9795\de$ & & & \\
$c$ & $x$ intercept of plane &  1961.810 & & & \\
$d$ & $u$ coordinate offset &  1227.920 & & & \\
$g$& $(a^2\sin^2\theta\,+\,b^2\cos^2\theta)^{1/2}$ &1700.726 & & &\\
$\gamma$ &$ 1-c^2\sin^2\theta/{g^2}$&  0.265634 & & & \\
$\alpha$ & semimajor axis ellipse&  869.665 & & &\\
$\beta$ & semiminor axis ellipse&  847.599 & & & \\
$\phi_1$& half-angle cone 1 &  $38.8953\de$ & & & \\
$\psi_1$& tilt cone 1 &  $-52.0529\de$ & & &\\
$\phi_2$ & half-angle cone 2&  $23.5993\de$ & & &\\
$\psi_2$ &tilt cone 2 &  $-30.1844\de$ & & &\\
$A_t$ &top of ellipse&  1721.997 & $-266.148$ & 0 & s\\
$A_t$ & &  8168.367 & $44.177$ & 0 & p\\
$A_b$ &bottom of ellipse &  557.696 & $-1558.305$ & 0 & s\\
$A_b$ & &  7386.922 & $-1509.725$ & 0 & p\\
$B$ & extrema of ellipse &  1139.847 & $-912.226$ & $\pm847.599$ & s\\
$B$ & &  7777.645 & $-732.774$ & $\pm847.599$ & p\\
$Q_{0}$& intersection of axes &  1552.150 & $-1242.2$ & 0 & s \\
$Q_{0}$& &  8262.490 & $-942.058$ & 0 & p \\
$Q_{1e}$& axis 1 intercept ellipse &  1222.935 & $-820.0142$ & 0 & s \\
$Q_{1e}$& &  7833.411 & $-621.883$ & 0 & p \\
$Q_{1s}$& axis 1 intercept spheroid &  1374.021 & $-1013.765$ & 0 & s\\
$Q_{1s}$& &  8030.327 & $-768.820 $ & 0 & p\\
$Q_{2e}$& axis 2 intercept ellipse &  1086.548 & $-971.378$ & 0 & s\\
$Q_{2e}$& &  7741.872 & $-803.908$ & 0 & p\\
$Q_{2s}$ & axis 2 intercept spheroid&  1300.929 & $-1096.073$ & 0 & s \\ 
$Q_{2s}$ & $=$ point $S$ in Equation 9&  7981.584 & $-867.517 $ & 0 & p \\ [1ex]
\hline
\end{tabular}
\end{center}
\vspace{0.2mm}
Parameters of SPT3G Secondary Mirror: Variables refer to definitions 
in the Appendix.
All dimensions in millimeters. The spheroid is the mirror
surface, the ellipse is the mirror edge. Subscript 1
refers to the cone between the prime focus and the secondary.
Subscript 2 refers to the cone between the secondary and
the Gregorian focus.
Coordinate system ``s" has its origin halfway between the two focii
and the $x$ axis is the line between the two focii.
Coordinate system ``p" is the global coordinate system with
the origin at the vertex of the primary.
\label{table:parameters}
\end{table}

\newpage
\begin{lstlisting}

/* file spt3gsecondary.c */
/* Calculate spt coordinate values from memo */
/* A. Stark */
/* 7-4-14 */
/* compile with  command
cc spt3gsecondary.c -lm -o spt3gsecondary
*/

#include <math.h>
#include <stdio.h>

        double fp;
	double theta_s;
        double xpe, ype;
        double theta_e;
        double theta_r;
        double theta_c;
        double theta_1,theta_2;
        double a, b, fs, f_0;
        double rtod;
	double i_p;
	double eslope;
	double xppe, yppe, zppe;
	double xdaggere, ydaggere, zdaggere;

main()
{
        double norm(), dot();
        void   stoptoprimary();
        void   ptou(),utop(),utopp(),pptou(),pptop(),ptopp();
        void   ppptop(), pptopppp();
	void   daggertou();
	double uonellipse(), pponellipse(), pponcone();
	double daggeronellipse(), daggeroncone();
	double quadp(), quadm();

        double yc,xc;
        double r2, k, e;
        double alpha, beta, gamma, nu, lambda, mu;
        double rho, sigma, tau;
        double F1x,F1y,F1z ;
        double Scx,Scy,Scz ;
        double F2x,F2y,F2z ;
        double f_1,f_2;
        double M;
        double i_s,  i_D;
        double iota;
        double temp;
        double xpptop, xppbot, ypptop, yppbot;
        double xdaggertop, xdaggerbot, ydaggertop, ydaggerbot;
        double omega;
        double upsilon;
        double ae, be;
        double J,K,L,N,P,Q;
	double p, q, r, s;
	double fparaxial;

        double x, y, z;
        double xp, yp, zp;
        double xpp, ypp, zpp;
        double xdagger, ydagger, zdagger;
        double xppp, yppp, zppp;
        double xpppp, ypppp, zpppp;
	double xppppmin, yppppmin, zppppmin;
	double Tslope, Bslope;
	double Atemp,Btemp,Ctemp;



        rtod = 45.0/atan(1.0);     /* convert radians to degrees */

        yc = 5300.0;               /* vertical offset of primary */
        fp = 7000.0;               /* focal length of primary */

        i_p = atan(yc/(2*fp))*rtod;  /* angle of incidence at primary */

        printf("\n i_p = %12.8gdeg\t angle of incidence at primary \n", i_p);

/* F1 is vector to prime focus */
        F1x = 7000.0; F1y = 0.0; F1z = 0.0;
        printf("\n F1 = %12.8gmm, %12.8gmm, %12.8gmm\t prime focus",
                 F1x,F1y,F1z);

/* slope of central ray between primary and secondary */
        alpha = (4.0*fp*yc)/(yc*yc - 4.0*fp*fp);
        printf("\nalpha = %12.8g\t slope of central ray", alpha);

/* Major and minor axes are defining values for SPT3G secondary */
	a = 1745.;
	b = 1644.556;
/* radius of curvature and conic constant of secondary */
/* these values, chosen here, define the shape of the secondary */

	e = sqrt(1.0 - b*b/(a*a));
	k = -1.0*e*e;
	r2 = a*(1.0+k);
	/*
        r2 = 1549.89366;
        k = -0.111809;
	e = sqrt(-k);
	*/
        printf("\n r2 = %12.8gmm\t secondary radius of curvature",r2);
        printf("\n k = %12.8g\t conic constant of secondary",k);
        printf("\n e = %12.8g\t eccentricity of secondary",e);

/* semi-major, semi-minor, focal distance of secondary */
	/*
        a = r2/(k+1.0);
        b = r2/sqrt(k+1.0);
	*/
        f_0 =  sqrt(a*a-b*b);
        fs = a - f_0;
        printf("\n a = %12.8gmm\t major axis of secondary spheroid",a);
        printf("\n b = %12.8gmm\t minor axis of secondary spheroid",b);
        printf("\n f_0 = %12.8gmm\t focal distance 0 of spheroid\n",
                   f_0);
        printf("\n fs = %12.8gmm\t vertex to f_1 distance of spheroid\n",
                   fs);


/* rotation of secondary axis */
/* this value is chosen so that Dragone angle (calculated below) is zero */
        theta_s = 15.323;
        printf("\n theta_s = %12.8gdeg\t rotation secondary axis",theta_s);

/* solve for intersection of central ray with secondary */
        rho = (cos(theta_s/rtod)+alpha*sin(theta_s/rtod))/a;
        sigma = (a - fs)/a;
        tau = (sin(theta_s/rtod)-alpha*cos(theta_s/rtod))/b;

        xc = fp + (-rho*sigma+sqrt(rho*rho*sigma*sigma -
              (rho*rho + tau*tau)*(sigma*sigma - 1.0)))/
              (rho*rho + tau*tau ) ;

        Scx = xc; Scy = alpha*(xc-fp); Scz=0.0;
        printf("\n C = %12.8gmm, %12.8gmm, %12.8gmm\t"
		"central ray at secondary", Scx,Scy,Scz);

/* F2 is vector to Gregorian focus */
        F2x = fp-2.0*(a-fs)*cos(theta_s/rtod); 
        F2y = -2.0*(a-fs)*sin(theta_s/rtod); 
        F2z =  0.0;
        printf("\n F2 = %12.8gmm, %12.8gmm, %12.8gmm\t Gregorian focus", 
                F2x,F2y,F2z);

/* distance from Gregorian focus to secondary along central ray */
        f_2 = norm(F2x-Scx,F2y-Scy,F2z-Scz);
        printf("\n f_2 = %12.8gmm\t Gregorian focus to secondary",f_2);
	f_2 = 2180.0;

/* distance from prime focus to secondary along central ray */
        f_1 = norm(F1x-Scx,F1y-Scy,F1z-Scz);
        printf("\n f_1 = %12.8gmm\t prime focus to secondary\n",f_1);
	f_1 = 1310.0;


/* magnification of secondary, is negative for Gregorians */
        M = -f_2/f_1;
        printf("\n M = %12.8g\tmagnification\n", M);

/* solve for angle of incidence of central ray at secondary */
        i_s = 0.5*acos(dot(F2x-Scx,F2y-Scy,F2z-Scz,
                           F1x-Scx,F1y-Scy,F1z-Scz)
                           /(f_2*f_1))*rtod;
        printf(" \n i_s = %12.8gdeg\t angle of"
		"incidence at secondary \n", i_s);

        printf("\n f_0 = %12.8g\t\n", f_0);
/*
	f_0 = 0.5*sqrt(f_1*f_1+f_2*f_2-2.0*f_1*f_2*cos(2.0*i_s/rtod));
        printf(" f_0 second time = %12.8g\t\n", f_0);
*/

/* solve for Dragone angle */
        i_D = atan((1.0 - M)*tan(i_s/rtod)+M*tan(i_p/rtod))*rtod;
        printf(" \n i_D = %12.3gdeg\t Dragone angle\n", i_D);

/* angle that rotates dagger coordinate system to p coordinate system */

        theta_r = -30.1844;
        printf("\n theta_r = %12.8gdeg\t ", theta_r);

        theta_r = (-1.0)* (theta_s + 2.0*i_p - 2.0*i_s);
        printf("\n alternate theta_r = %12.8gdeg\t ", theta_r);

        theta_1 = rtod * asin((((b*b)/(a*f_1) - 1.0)/e));
        printf("\n theta_1 = %12.8gdeg\t rotate pp to p ", theta_1);

        theta_1 = rtod * 
		asin((f_2*f_2-f_1*f_1-4.0*f_0*f_0)/(4.0*f_0*f_1));
        printf("\n theta_1 = %12.8gdeg\t rotate pp to p ", theta_1);

	temp = f_1 * sin(theta_1/rtod) + f_0;
	theta_2 = rtod * atan(b*temp/(a*a*sqrt(1.0 - temp*temp/(a*a))));
        printf("\n theta_2 = %12.8gdeg ", theta_2);
	
	e = 2*f_0/(f_1+f_2);
        printf("\n e = %12.8g\t eccentricity ", e);

	p = 1.0 - e*e * sin(theta_2/rtod) * sin(theta_2/rtod);
        printf("\n p = %12.8g ", p);

	q = e*e * sin(2.0*theta_2/rtod);
        printf("\n q = %12.8g ", q);

	r = f_1*(cos(i_s/rtod)+e*sin(theta_2/rtod));
        printf("\n r = %12.8g ", r);

	r = f_2*(cos(i_s/rtod) - e*sin(theta_2/rtod));
        printf("\n r = %12.8g ", r);

	s = 1.0 - e*e * cos(theta_2/rtod) * cos(theta_2/rtod);
        printf("\n s = %12.8g ", s);

	fparaxial = r / (2.0 * cos(i_s/rtod));
        printf("\n fparaxial  = %12.8g\t central radius = %12.8g\n",
		fparaxial, 2.0*fparaxial);

        printf("\n Is S on ellipse? %12.8g ",uonellipse(Scx,Scy,Scz));

	theta_c= 23.5993;
        printf("\n theta_c = %12.8gdeg\t defined"
		"half angle of cone  ", theta_c);

	eslope = (-1.0*f_0*cos(theta_c)
		-a*cos(theta_r/rtod))/(a*sin(theta_r/rtod));
        printf("\n e slope Equation 44 = %12.8g ", eslope);

	theta_e = rtod * atan(eslope);
        printf("\t theta_e = %12.8gdeg\t angle of ellipse"
		"in dagger \n", theta_e);

	iota = (b*b*cos(theta_c/rtod))/(a*sin(theta_r/rtod));
        printf("\n iota Equation45 = %12.8g mm \n", iota);

	Tslope = tan((theta_r+theta_c)/rtod);
	Bslope = tan((theta_r-theta_c)/rtod);
        printf("\n Tslope = %12.8g  \n", Tslope);
        printf("\n Bslope = %12.8g  \n", Bslope);
	Atemp = a*a*Tslope*Tslope+b*b;
	Btemp = -2.0*b*b*f_0;
	Ctemp = f_0*f_0*b*b-a*a*b*b;
        if (quadp(Atemp,Btemp,Ctemp)>quadm(Atemp,Btemp,Ctemp)) {
		xdaggertop=quadp(Atemp,Btemp,Ctemp);
		printf("xdaggertop positive");
		} else {
		xdaggertop=quadm(Atemp,Btemp,Ctemp);
		printf("xdaggertop negative");
		}
	ydaggertop = Tslope*xdaggertop;
        printf("\n xdaggertop,ydaggertop= %12.8g mm, %12.8g mm ", 
		xdaggertop,ydaggertop);
        printf("\n Is top on ellipse pp? %12.8g ",
		daggeronellipse(xdaggertop,ydaggertop,0.0));
        printf("\n Is top on cone pp? %12.8g \n",
		daggeroncone(xdaggertop,ydaggertop,0.0));

	Atemp = a*a*Bslope*Bslope+b*b;
        if (quadp(Atemp,Btemp,Ctemp)*Bslope < 
		quadm(Atemp,Btemp,Ctemp)*Bslope) {
		xdaggerbot=quadp(Atemp,Btemp,Ctemp);
		printf("xxpbot positive");
		} else {
		xdaggerbot=quadm(Atemp,Btemp,Ctemp);
		printf("xxpbot negative");
		}
	ydaggerbot = Bslope*xdaggerbot;
        printf("\n xdaggerbot,ydaggerbot= %12.8g mm, %12.8g mm ", 
		xdaggerbot,ydaggerbot);
        printf("\n Is bot on ellipse dagger? %12.8g ",
		daggeronellipse(xdaggerbot,ydaggerbot,0.0));
        printf("\n Is bot on cone dagger? %12.8g ",
		daggeroncone(xdaggerbot,ydaggerbot,0.0));
	 
	/*
	xdaggertop = (a*b*b/cos((theta_c-theta_r)/rtod) - b*b*f_0)
		/(a*a*Tslope*Tslope + b*b);
	ydaggertop = Tslope*xpptop;
        printf("\n alternate xdaggertop,ydaggertop=%12.8g mm, %12.8g mm", 
		xdaggertop,ydaggertop);

	xdaggerbot = (a*b*b/cos((-1.0*theta_c-theta_r)/rtod) 
		- b*b*f_0)/(a*a*Bslope*Bslope + b*b);
	ydaggerbot = Bslope*xdaggerbot;
        printf("\n alternate xppbot,yppbot= %12.8g mm, %12.8g mm ", 
		xdaggerbot,ydaggerbot);
	*/
	temp = xdaggertop-xdaggerbot;
	ae = temp*temp;
	temp = ydaggertop-ydaggerbot;
	ae += temp*temp;
	ae = sqrt(ae)/2.0;
        printf("\n ae= %12.8g mm ", ae);

	xdaggere = (xdaggertop+xdaggerbot)/2.0;
	ydaggere = (ydaggertop+ydaggerbot)/2.0;
	zdaggere = b*sqrt(1.0 - (xdaggere-f_0)*(xdaggere-f_0)/(a*a) 
		- ydaggere*ydaggere/(b*b));
        printf("\n xdaggere,ydaggere,zdaggere="
		"%12.8g mm, %12.8g mm, %12.8g mm\t \n", 
		xdaggere,ydaggere,zdaggere);
        printf("\n Is daggere on ellipse dagger? %12.8g ",
		daggeronellipse(xdaggere,ydaggere,zdaggere));
        printf("\n Is daggere on cone dagger? %12.8g \n",
		daggeroncone(xdaggere,ydaggere,zdaggere));

	
	eslope = (ydaggertop-ydaggerbot)/(xdaggertop-xdaggerbot);
        printf("\n e slope= %12.8g ", eslope);

	theta_e = rtod * atan(eslope);
        printf("\t theta_e = %12.8gdeg\t angle of ellipse in dagger \n", 
		theta_e);

        printf("\n theta_e+theta_s = %12.8gdeg", theta_e + theta_s);
	
	iota = ydaggertop - eslope* xdaggertop;
        printf("\n iota = %12.8g mm", iota);
        printf("\n ydaggerbot = %12.8g mm", xdaggerbot*eslope+iota);


	daggertou(xdaggertop,ydaggertop,0.0,&x,&y,&z);
        printf("\n top of secondary edge = %12.8gmm, %12.8gmm, %12.8gmm", 
                x,y,z);
        printf("\n Is top on ellipse? %12.8g \n",uonellipse(x,y,z));

	daggertou(xdaggerbot,ydaggerbot,0.0,&x,&y,&z);
        printf("\n bottom of secondary edge = %12.8gmm, %12.8gmm, %12.8gmm", 
                x,y,z);
        printf("\n Is bottom on ellipse? %12.8g \n",uonellipse(x,y,z));

	daggertou(xdaggere,ydaggere,zdaggere,&x,&y,&z);
        printf("\n center of secondary edge = "
		"%12.8gmm, %12.8gmm, %12.8gmm\t \n", x,y,z);

        printf("\n Is center edge on ellipse? %12.8g \n",uonellipse(x,y,z));
	

        printf("\nInput mirror coordinate xpppp:  ");
        scanf("%lg",&xpppp);
        printf("Input mirror coordinate ypppp:  ");
        scanf("%lg",&ypppp);

	J = -1.0*(f_0*f_0*cos(theta_e/rtod)*cos(theta_e/rtod)+b*b);
        printf("\n J = %12.8g mm\t \n", J);

	K = -1.0*b*b;
        printf("\n K = %12.8g mm\t \n", K);

	temp = ydaggere*sin(theta_e/rtod)+(xdaggere-f_0)*cos(theta_e/rtod);
	L = -2.0*b*b*temp;
        printf("\n L = %12.8g mm\t \n", L);

	N = b*b*(b*b+f_0*f_0*cos(theta_e/rtod)*
		cos(theta_e/rtod)-temp*temp);
        printf("\n N = %12.8g mm\t \n", N);

	P = f_0*f_0*cos(theta_e/rtod)*sin(theta_e/rtod);
        printf("\n P = %12.8g mm\t \n", P);

	Q = a*a*ydaggere*cos(theta_e/rtod)
		-b*b*(xdaggere-f_0)*sin(theta_e/rtod);
        printf("\n Q = %12.8g mm\t \n", Q);

	zpppp = (P*xpppp+Q+a*sqrt(J*ypppp*ypppp
		+K*xpppp*xpppp+L*xpppp+N))/J;
        printf(" mirror sag zpppp = %12.8g\t \n", zpppp);

	xppppmin = (a*a*K*L-L*P*P-P*sqrt((L*L-4.0*K*N)*P*P+
			4.0*a*a*K*K*N-a*a*K*L*L))
			/(2.0*(K*P*P-a*a*K*K));

        xpppp = xppppmin; 
	ypppp = 0.0;
	zpppp = (P*xpppp+Q+a*sqrt(J*ypppp*ypppp
		+K*xpppp*xpppp+L*xpppp+N))/J;
        printf("\n lowest point on mirror="
		"%12.8gmm, %12.8gmm, %12.8gmm\t \n", xpppp,ypppp,zpppp);

	utopp(7981.579,-867.52618,0,&x, &y, &z);
	pptopppp(x,y,z,&xpppp, &ypppp, &zpppp);
        printf("\n central ray at mirror=" 
		"%12.8gmm, %12.8gmm, %12.8gmm\t \n", 
			xpppp, ypppp, zpppp);


        printf("\n");

}

double norm(x,y,z)      /* norm of a vector */
double x, y, z;
{
        return(sqrt(x*x+y*y+z*z));
}

double dot(x1,y1,z1,x2,y2,z2)       /* dot product of two vectors */
double x1,y1,z1,x2,y2,z2;
{
        return(x1*x2+y1*y2+z1*z2);
}

double uonellipse(double xonel, double yonel, double zonel)
/* Is the point on the spheroid in unprimed coords? */
{
	double temponel, sumonel;

	temponel = ((xonel-fp)*cos(theta_s/rtod)
		+yonel*sin(theta_s/rtod)+f_0)/a ;
       	sumonel = temponel*temponel;
	temponel = ((xonel-fp)*sin(theta_s/rtod)
		-yonel*cos(theta_s/rtod))/b;
	sumonel += temponel*temponel;
	temponel = zonel/b;
	sumonel += temponel*temponel;
	return(sumonel);
}

double pponellipse(double xonel, double yonel, double zonel)
/* Is the point on the spheroid in double prime coords? */
{
	double temponel, sumonel;

	temponel = (xonel+f_0)/a;
       	sumonel = temponel*temponel;
	temponel = yonel/b;
	sumonel += temponel*temponel;
	temponel = zonel/b;
	sumonel += temponel*temponel;
	return(sumonel);
}

double daggeronellipse(double xonel, double yonel, double zonel)
/* Is the point on the spheroid in dagger coords? */
{
	double temponel, sumonel;

	temponel = (xonel-f_0)/a;
       	sumonel = temponel*temponel;
	temponel = yonel/b;
	sumonel += temponel*temponel;
	temponel = zonel/b;
	sumonel += temponel*temponel;
	return(sumonel);
}

double daggeroncone(double xonco, double yonco, double zonco)
/* Is the point on the cone in dagger coords? */
{
	double temponco;

	temponco = sqrt(xonco*xonco+yonco*yonco+zonco*zonco)
		*cos(theta_c/rtod);
	return((xonco*cos(theta_r/rtod)+yonco
		*sin(theta_r/rtod))/temponco);
}

void ptou(double xp, double yp, double zp, 
	double *x, double *y, double *z)
/* convert primed to unprimed coordinates */
{
       *x=cos(2.0*i_p/rtod)*(xp)+sin(2.0*i_p/rtod)*yp+fp;
       *y=(-1.0)*sin(2.0*i_p/rtod)*(xp)+cos(2.0*i_p/rtod)*yp;
       *z=zp;
}

void ptopp(double xp, double yp, double zp, 
	double *xpp, double *ypp, double *zpp)
/* convert primed to double primed coordinates */
{
       *xpp= cos(theta_1/rtod)*(xp) - sin(theta_1/rtod)*zp;
       *ypp=yp;
       *zpp= sin(theta_1/rtod)*(xp)+cos(theta_1/rtod)*zp;
}

void pptou(double xpp, double ypp, double zpp, 
	double *x, double *y, double *z)
/* convert double primed to unprimed coordinates */
{
       *x=cos(theta_s/rtod)*(xpp)-sin(theta_s/rtod)*ypp+fp;
       *y=sin(theta_s/rtod)*(xpp)+cos(theta_s/rtod)*ypp;
       *z=zpp;
}

void daggertou(double xdagger, double ydagger, 
	double zdagger, double *x, double *y, double *z)
/* convert double primed to unprimed coordinates */
{
       *x=cos(theta_s/rtod)*(xdagger - 2.0*f_0)
       		-sin(theta_s/rtod)*ydagger +fp;
       *y=sin(theta_s/rtod)*(xdagger - 2.0*f_0)
       		+cos(theta_s/rtod)*ydagger;
       *z=zdagger;
}

void pptop(double xpp, double ypp, double zpp, 
	double *xp, double *yp, double *zp)
/* convert double primed to primed coordinates */
{
       *xp= cos(theta_1/rtod)*(xpp)+sin(theta_1/rtod)*zpp;
       *yp=ypp;
       *zp= (-1.0)*sin(theta_1/rtod)*(xpp)+cos(theta_1/rtod)*zpp;
}

void utopp(double x, double y, double z, 
	double *xpp, double *ypp, double *zpp)
/* convert unprimed to double primed coordinates */
{
       *xpp=cos(theta_s/rtod)*(x - fp) + sin(theta_s/rtod)*y;
       *ypp= (-1.0)* sin(theta_s/rtod)*(x - fp)+cos(theta_s/rtod)*y;
       *zpp=z;
}

void utop(double x, double y, double z, 
	double *xp, double *yp, double *zp)
/* convert unprimed to primed coordinates */
{
       *xp=cos(2.0*i_p/rtod)*(x - fp) - sin(2.0*i_p/rtod)*y;
       *yp= sin(2.0*i_p/rtod)*(x - fp)+cos(2.0*i_p/rtod)*y;
       *zp=z;
}

void pptopppp(double xpp, double ypp, double zpp, 
	double *x, double *y, double *z)
/* convert double primed to quadruple primed coordinates */
{
       *x=cos(theta_e/rtod)*(xpp-xppe)+sin(theta_e/rtod)*(ypp-yppe);
       *y=(-1.0)*zpp;
       *z=(-1.0)*sin(theta_e/rtod)*(xpp-xppe)+cos(theta_e/rtod)*(ypp-yppe);
}

void stoptoprimary(double xs, double ys, double zs, 
	double *x, double *y, double *z)
/* project points at the stop onto primary */
{
       *y = 2.0*fp*ys*(xs-fp-sqrt((xs-fp)*(xs-fp)
               +ys*ys+zs*zs))/(ys*ys+zs*zs);
       *x = (*y)*(xs-fp)/ys+fp;
       *z = (*y)*zs/ys;
}

void ppptop(double xppp, double yppp, double zppp, 
	double *xp, double *yp, double *zp)
/* convert mirror coordinates to primed coordinates */
{
       *xp=cos(theta_e/rtod)*xppp-sin(theta_e/rtod)*zppp+xpe;
       *yp=sin(theta_e/rtod)*xppp+cos(theta_e/rtod)*zppp+ype;
       *zp=-yppp;
}

double quadp(double A, double B, double C)
{
	double surd;

	surd = B*B-4.0*A*C;
	if (surd < 0) {
		printf("\nbad surd in quadp\n");
		return(0.0);
        } else {
		return((-B+sqrt(surd))/(2.0*A));
        }
}

double quadm(double A, double B, double C)
{
	double surd;

	surd = B*B-4.0*A*C;
	if (surd < 0) {
		printf("\nbad surd in quadm\n");
		return(0.0);
        } else {
		return((-B-sqrt(surd))/(2.0*A));
        }
}
\end{lstlisting}

\newpage

\end{document}